\documentstyle[12pt,aasms4]{article}

\begin{document}

\title{Historic Light Curve and Long-term Optical Variation of BL Lacertae 2200+420}

\author{J.H. Fan}

\affil{CRAL Observatoire de Lyon,9, Avenue Charles Andre, 69 563 Saint-Genis-Laval Cedex, France, e-mail: jhfan@zorglub.univ-lyon1.fr\\
Center for Astrophysics, Guangzhou Normal University, Guangzhou 510400, China, e-mail: 
jhfan@guangztc.edu.cn \\
Joint Laboratory for Optical Astronomy, Chinese Academy of Sciences, China} 
\and
\author{G.Z. Xie} 
\affil{Yunnan Observatory, Chinese Academy of Sciences, Kunming 650011,China }
\and
\author{E. Pecontal, A. Pecontal and Y. Copin}

\affil{CRAL Observatoire de Lyon,9, Avenue Charles Andre, 69 563 Saint-Genis-Laval Cedex, France}

\begin{abstract}
In this paper,  historical optical(UBVRI) data and newly observed data from the Yunnan Observatory of China( about100 years)
are presented for BL Lacertae.  Large variations of $\Delta U = 5^m.12 $, $\Delta B = 5^m.31$, 
$\Delta V = 4^m.73 $, $\Delta R = 2^m.59 $, $\Delta I = 2^m.54 $,   and 
 color indices of $ U-B = -0.11 \pm 0.20$, $B-V= 1.0 \pm 0.11$,
$V-R= 0.73\pm 0.19$,$V-I= 1.42\pm 0.25$,$R-I= 0.82\pm 0.11$,and $B-I= 2.44\pm 0.29$ have been obtained from the literature;
The Jurkevich method is used to investigate the existence of periods in the B band light curve, and a long-term period of  
14 years is found. The 0.6 and 0.88 year periods reported by Webb et al.(1988) are confirmed. 
In addition, a close relation between B-I and B is found, suggesting that the spectra flattens
when the source brightens. 
\end{abstract}


\section{Introduction}

The nature of active galactic nuclei (AGNs) is still an open problem, the study of variability about AGNs, 
such as periodicity analysis of light curves, can yield valuable information about the nature of AGNs and the 
implications for quasar modeling are extremely important(Blandford, 1996).  In AGNs, the long-term optical variation has in some cases 
claimed to be periodic.  For example, the 15-year period found in the light curve of 3C120 (Belokon 1987;Hagen-thorn et al. 1997),
 the 13.6-year period found in the light curve of ON231(Liu et al. 1995),  the 14.2-year period found in the light curve of PKS 0735+178
 (Fan et al. 1997), and the 14.0-year period found in the light curve of NGC4151 (Fan \& Su 1997).  The 11.4-year period found in the 
light curve of 3C345 (Webb et al. 1988) predicted the 1991 outburst (Kidger \& Takalo 1990); The 12-year period found in 
the light curve of OJ 287 (Sillanpaa et al. 1988a; Kidger et al. 1992) successfully predicted the optical outburst in the fall of 1994
 (Sillanpaa et al. 1996a,b; Arimoto et al. 1997). 

Photometric observations of AGNs are important to construct their light curves and to study their variation behavior on
different time scales.  Using the 1-m telescope at Yunnan Observatory, the 1.56-m telescope at Shanghai Observatory, and the 2.16-m 
telescope at Beijing Observatory, we have monitored dozens of AGNs, including BL Lac objects and Seyfert galaxies (Xie et al. 1987, 
1988a,b, 1990, 1992, 1994; Fan et al. 1997, Bai et al. 1998).

BL Lacertae (2200+420), the archetype of its class, lies in a giant elliptical galaxy at a redshift of $\sim 0.07$ (Miller et al. 1978).
It is one of the best-studied objects at optical and radio bands.  Superluminal components have been observed from the source(Mutel
 \& Phillips 1987; Vermeulen \& Cohen 1994; Fan et al. 1996).  Its optical and radio emissions are both  variable and polarized, and its
radio and optical polarizations are correlated (Sitko et al. 1985).  Recently, observations  with EGRET on the Compton Gamma Ray
 Observatory (CGRO) between 1995 January 24 and 1995 February 14 indicate a flux of $(40\pm 12)\times 10^{ -8} photon/cm^{2} /s$ above $100
MeV$ (Catanese et al. 1997), but there is no evidence of gamma rays in Whipple observations (Kerrick et al. 1995; Quinn et al. 1995).  It
has been observed in the optical for about 100 years, with early observations presented in the papers of Shen \& Usher (1970) and
Webb et al. (1988).  It has been observed for about 14 years in our own monitoring program.

In this paper, we present BL Lacertae  historical optical data and new observations in UBVRI bands, and discuss its long-term variation.
  The paper has been arranged as follows: In section 2, we present the data; in section 3, the periodicity analysis; 
and in section 4, the discussion.

\section{Observations}
\subsection{Data}

During 1995 and 1996, BL Lacertae was observed with the 1-m RRC telescope at Yunnan Observatory which is equipped with a direct CCD
camera at the Cassegrain focus.  The filters are standard Johnston broadband filters.  The standard stars given in the paper of Smith
 et al.(1985) were used.  The daily averaged magnitudes obtained from observations are presented in table 1.  All the detailed magnitudes
  will appear in a separate paper with other BL Lac objects data obtained during the same period (Bai, et 
al. 1998). 

\begin{table}
\caption[]{Daily Averaged Magnitudes of BL Lacertae}
\begin{tabular}{ccccc}
\hline\noalign{\smallskip}
Date       &  B          &  V          &  R            &  I \\ 
\noalign{\smallskip}
\hline\noalign{\smallskip}
95-10-21   &             & 16.32(0.09) &               &             \\ 
95-10-22   &             & 15.66(0.10) &  14.87(0.02)  & 14.23(0.05) \\ 
95-10-28   & 15.88(0.02) & 15.31(0.12) &  14.69(0.05)  & 13.88(0.08) \\ 
95-10-29   & 16.09(0.15) & 15.29(0.08) &  14.67(0.04)  & 13.86(0.07) \\ 
95-12-22   & 16.05(0.06) & 15.37(0.09) &  14.48(0.05)  & 13.76(0.03) \\ 
95-12-23   &             &             &  14.58(0.05)  & 13.89(0.05) \\ 
96-08-14   &             &             &  14.77(0.07)  & 14.00(0.07) \\ 
96-08-15   &             &             &  14.71(0.08)  & 14.06(0.07) \\ 
96-08-23   &             & 15.05(0.05) &  14.46(0.04)  & 13.84(0.05) \\ 
96-08-24   & 15.80(0.10) & 15.20(0.05) &  14.56(0.06)  & 13.58(0.05) \\ 
96-08-25   & 16.00(0.10) & 15.38(0.05) &  14.68(0.07)  & 13.85(0.06) \\ 
\noalign{\smallskip}
\hline    
\end{tabular}
\end{table}

The historic data are from the literature (Bertaud et al. 1969; Du Puy et al. 1969; Racine 1970; Cannon et al. 1971; Visvanathan 1973; 
Bertaud et al. 1973;Barbier et al. 1978; Miller et al. 1978; Miller \& McGimsey 1978; O'Dell et al. 1978;Puschell \& Stein 1980; 
Barbieri et al. 1982; Sitko et al. 1983, 1985; Hagen-Thorn et al. 1984; Moles et al. 1985; Corso et al. 1986; Smith et al. 1987; 
Kidger 1988; Sillanpaa et al; 1988b, 1991; Webb et al. 1988; Bregman et al; 1990; Mead et al. 1990; Kawai et al. 1991; Sitko \& Sitko 1991;
 Takalo 1991; Carini et al. 1992; Okyudo 1993; Qogun et al. 1994; Corbett et al. 1996a; Maesano et al. 1997)
 with the $m_{pg}$ (Qogun  et al. 1994) changed into B by B-$m_{pg}$=0.28 (Kidger 1989; Lu 1972).  Also those observed by our group 
(Xie et al. 1988b, 1990, 1992, 1994) and those presented in table 1 are used in our analysis.  In order to discuss the long-term 
variability, we have included data estimated from Figure 1 in the paper of Shen \& Usher (1970) with relatively large data 
uncertainties (less than one month). The data in UVRI bands are shown in Figure 1a,b,c,d.  B band data are shown in Figure 2.

\begin{figure}
\epsfxsize=12cm
$$
\epsfbox{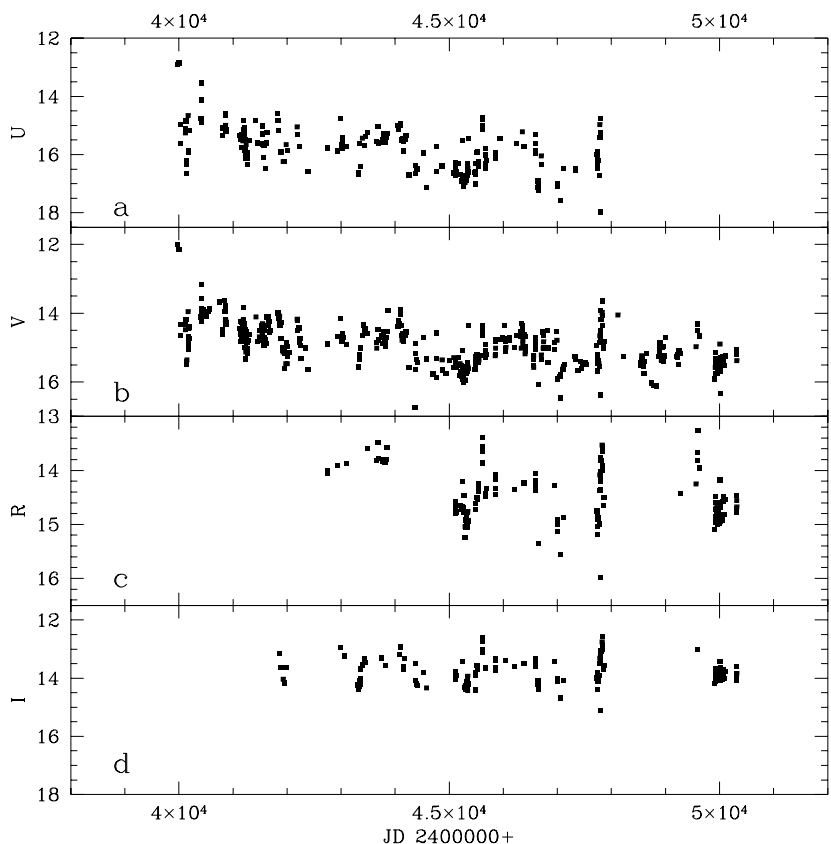}
$$
\caption{(a) The long-term U light curve of BL Lacertae from 1968 to 1989, there are no new data
 avilable from the litterature after 1989. 
(b) The long-term V light curve of BL Lacertae from 1968 to 1996.
(c) The long-term R light curve of BL Lacertae from 1975 to 1996, the discontinuity of the light curve between 
2447860 and 2449273 is due to the lack  of observations in this band
(d) The long-term I light curve of BL Lacertae from 1973 to 1996, the discontinuity of the light curve between 2447860 
and 2449599 is due to the lack of observations in this band}
\end{figure}

\begin{figure}
\epsfxsize=12cm
$$
\epsfbox{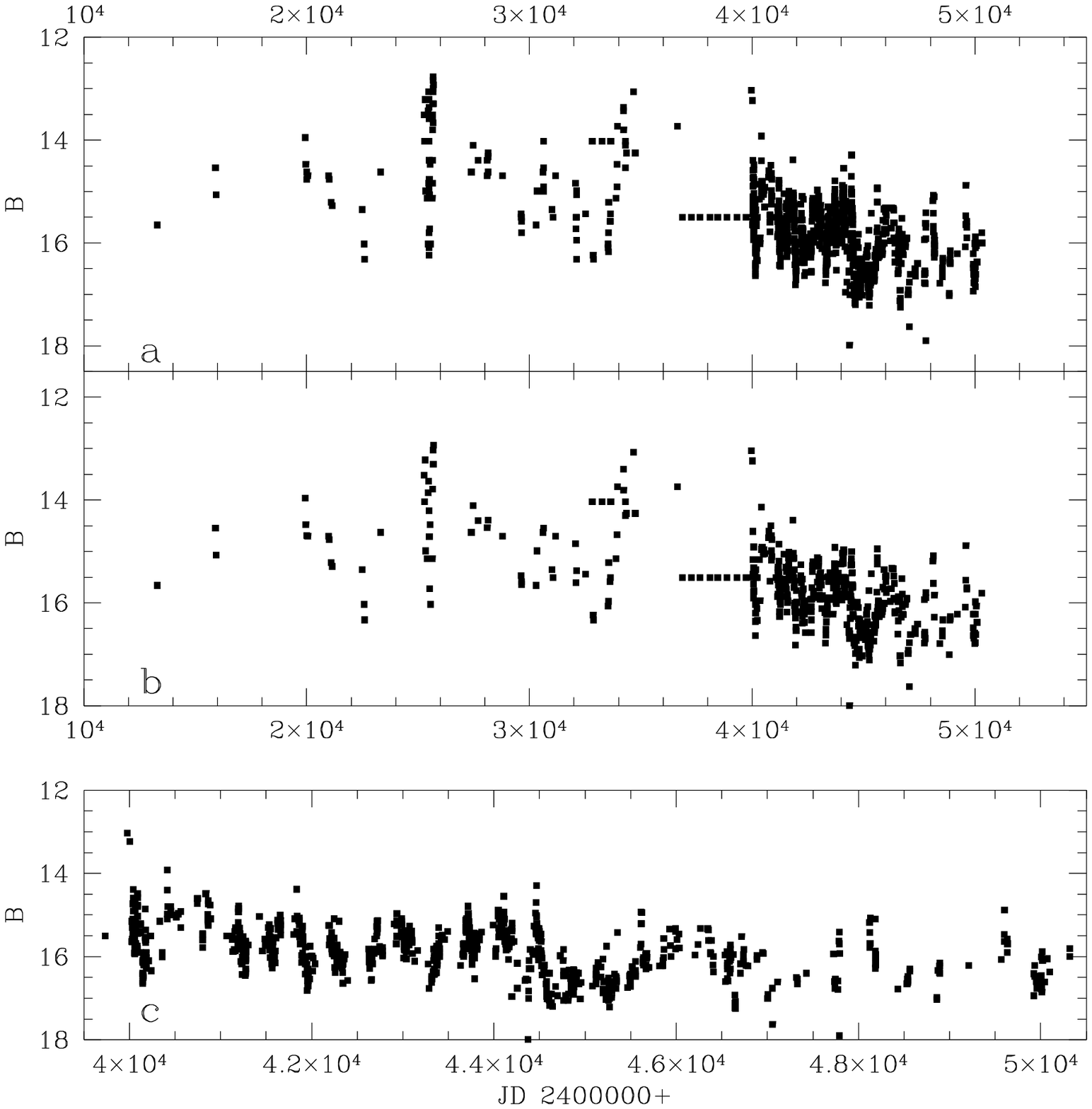}
$$
\caption{(a) The long-term B light curve of BL Lacertae covering the period of 1896 - 1996.
(b) The 5-day-averaged long-term B light curve of BL Lacertae  covering the period of 1896 - 1996
(c) The long-term B light curve of BL Lacertae covering the period of 1970 - 1996 }
\end{figure}

\subsection{Variations}

From the literature, we find that the largest variations in the five bands are $\Delta U = 5^m.12 (17.97-12.85)$,
 $\Delta B = 5^m.31  (17.99-12.68)$,$\Delta V = 4^m.73 (16.73-12.00)$, $\Delta R = 2^m.73  (15.99-13.26)$,
 $\Delta I = 2^m.54 (15.10-12.56)$. If we only consider the data observed in the same period for the five bands, we
find the large variations are respectively:$\Delta U = 3^m.39 (17.97-14.58)$,
 $\Delta B = 3^m.70  (17.99-14.29)$,$\Delta V = 3^m.11 (16.73-13.62)$, $\Delta R = 2^m.73(15.99-13.26)$,
 $\Delta I = 2^m.54 (15.10-12.56)$. The variations suggest that the largest variation increases with decreasing  wavelength. 
 In 1995/96, BL Lacertae was found in a low state.  It reached V = 15.9--15.7 in June 1995 (Maesano et al. 1997 and reference therein)
 with emission lines and absorption features being observed (Vermeulen et al. 1995).  Besides the broad $H_\alpha$ emission line,
 Corbett et al. (1996a,b) also identified narrow lines of $[N_{II}]\lambda 6583$, $[O_I]\lambda 6300$, $[Fe_{VII}]\lambda 6087$. 
 In our own monitoring program, 1995 observations are consistent with those of Maesano et al.(1997).

\begin{figure}
\epsfxsize=12cm
$$
\epsfbox{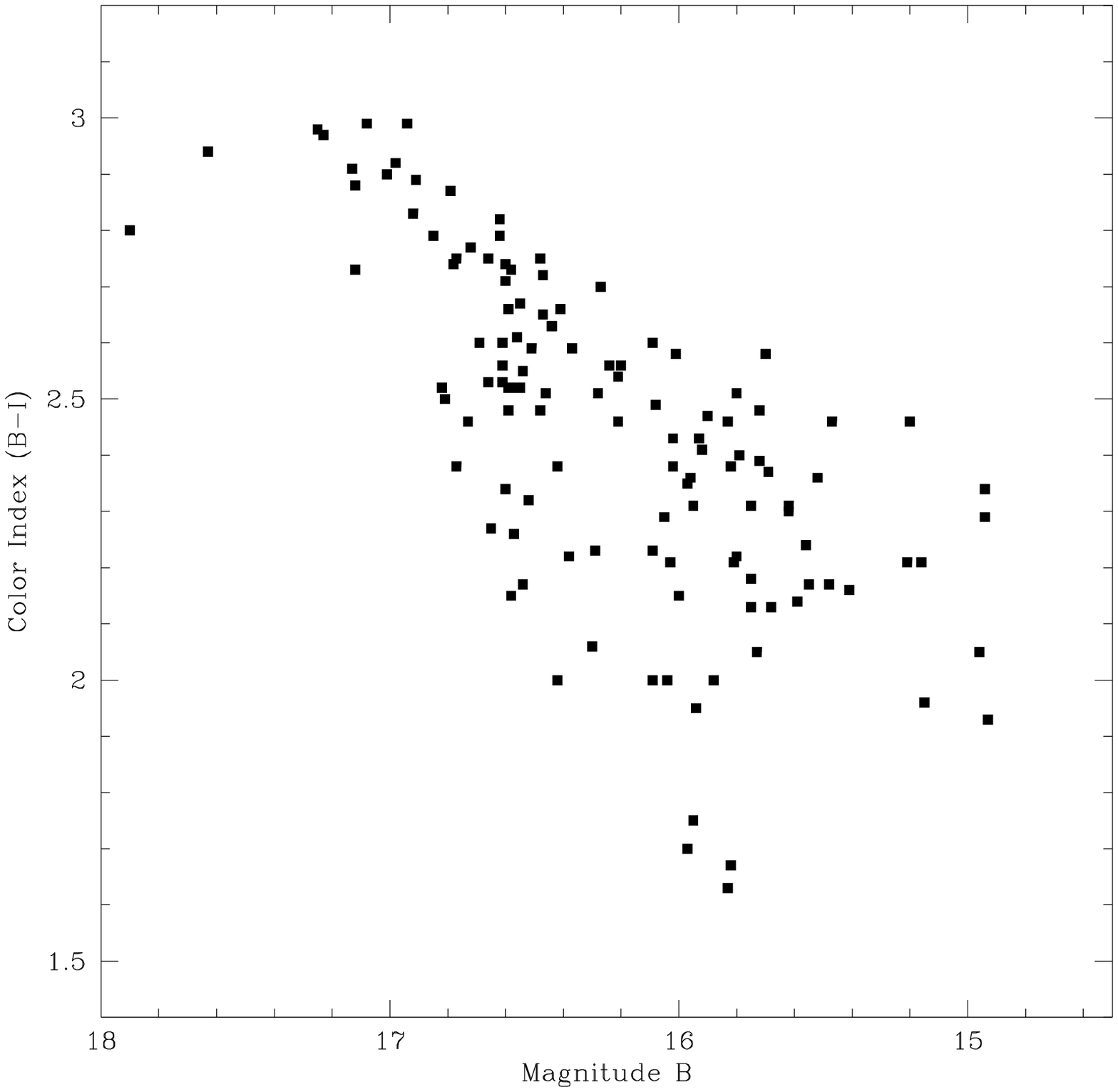}
$$
\caption{ Plot of B-I against B, showing that the spectrum changes with 
the brightness of the source }
\end{figure}

From the data, it can be seen that there is no correlation between B-V and U-B, or between V-R and R-I.  A correlation has been found
 for B-I and B (see Figure 3):  $B=(1.31\pm0.02)(B-I)+(13.03 \pm 0.1)$ with a correlation coefficient $r=0.67 $. The probability 
of the relationship having occured by chance is much less than 1$\%$. For the B and I data pairs, the I data change in the range of
12.60 to 15.10 while the B data in the range of 14.93 to 17.90. This 
correlation means that the spectrum flattens when the source brightens and the spectrum steepens when the source dims. But no
 correlation is found  for U-B and V, B-V and V, R-I and I, or V-I and I.  The color indices, B-V and R-I show narrower
 dispersions than other color indices (see Figure 4): $U-B=-0.11\pm0.20$; $V-R= 0.73\pm0.19$; $V-I=1.42\pm0.25$; $B-V=1.00\pm0.11$;
 $R-I=0.82\pm0.11$. $B-I=2.44\pm0.29$.

\begin{figure}
\epsfxsize=12cm
$$
\epsfbox{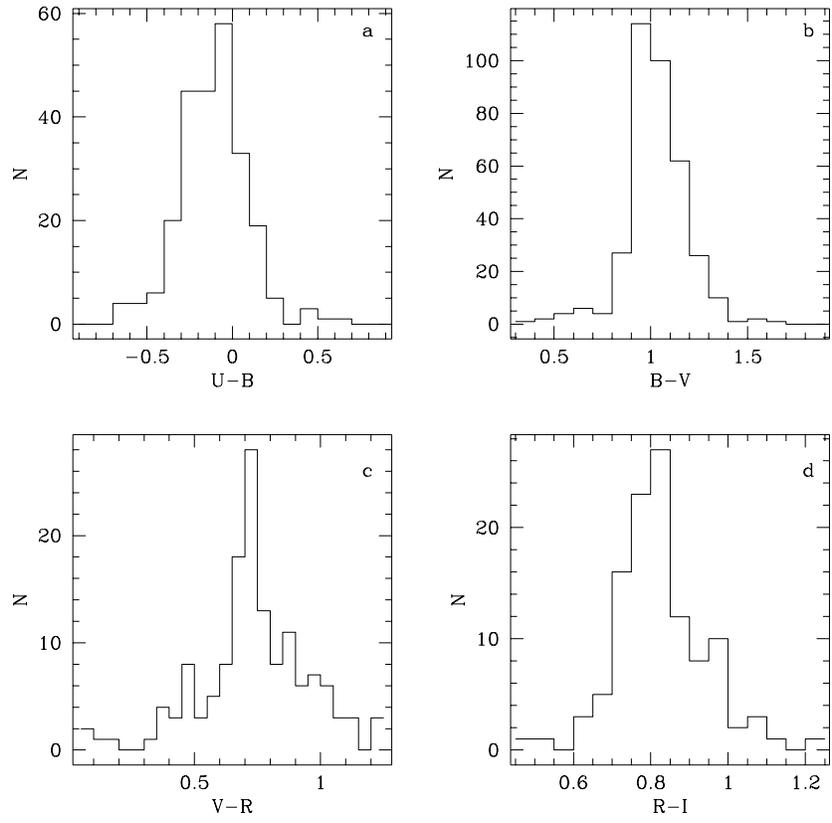}
$$
\caption{(a) Histogram of U-B color index;
(b) Histogram of B-V color index;
(c) Histogram of V-R color index ;
(d) Histogram of R-I color index  }
\end{figure}

\section{Jurkevich's Analysis Method}

 We will use the Jurkevich method for the periodicity analysis in the light curve of BL Lacertae. The Jurkevich method(Jurkevich 1971) is based on the expected mean square deviation and it is less inclined to generate spurious
 periodicity than the Fourier analysis.  It tests a run of trial periods around which the data are folded.  All data are assigned 
to $m$ groups according to their phases around each trial period. The variance $V_i^2$ for each group and the sum $V_m^2$ of all groups
 are computed.  If a trial period equals  the true one, then $V_m^2$ reaches its minimum. So, a ``good'' period will give a much reduced
 variance relative to those given by other false trial periods and with almost constant values.  The computation of the variances has
 been described in the paper of Jurkevich et al.(1970).  Kidger et al.(1992) introduced a  fraction of the variance
 $$f=\frac{1-V_m^2}{V_m^2}$$, where $V_m^2$ is the normalized value.
  In the normalized plot, a value of $V_m^2 =1$
 means $f=0$ and hence there is no periodicity at all.  The best periods can be identified from the plot: a value of $f\geq 0.5$ suggests
 that there is a very strong periodicity and a value of $f < 0.25$ suggests that the periodicity, if genuine, is a weak one. A further test
 is the relationship between the depth of the minimum and the noise in the ``flat'' section of the $V_m^2$ curve close to the adopted
 period.
  If the absolute value of the relative change of the minimum to the ``flat'' section is large enough as compared with the standard error
 of this ``flat'' section, the periodicity in the data can be considered as significant and the minimum as highly
 reliable(Kidger et al. 1992).  Here we consider the half width at half minimum as the ``formal'' error as did Jurkevich (1971).

\begin{figure}
\epsfxsize=12cm
$$
\epsfbox{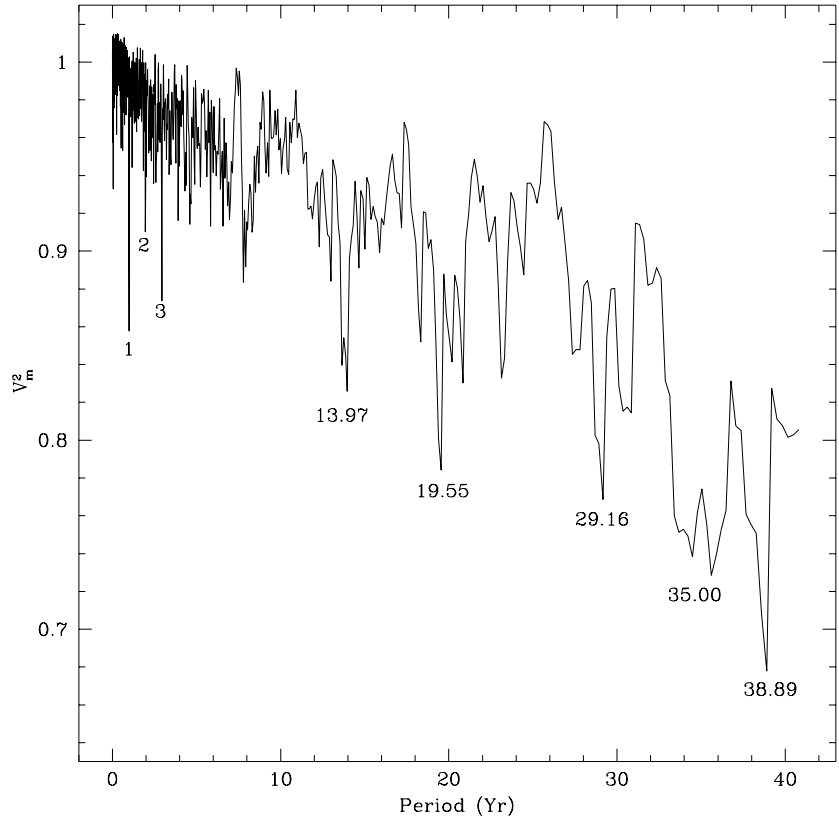}
$$
\caption{ Diagram of $V_{m}^2$ against the trial periods for the 5-day-averaged data covering the observation period
  1896 - 1996 ($m$=10 is adopted). The deepest minimum corresponds to a period of $38.89$ years, other minima to periods
 of $13.97$, $19.55$, $29.16$, $35.00$ years respectively.  Periods of $1$, $2$, and $3$ years are also
 conspicuous as compared with the ``flat'' noise }
\end{figure}

We use the B-band photometry to investigate the long-term period of variation since most data were obtained in this filter. In order to 
avoid the problem that sampling was more frequent during recent decades, we averaged the photometry into 5-day intervals (Figure 2b).  This 
is short enough as compared to the long-term period (years) and thus unlikely to distort the long-term variation too much.  When 
the Jurkevich method is applied to the averaged data, periods of 13.97, 19.55, 29.16, 35.00 and 38.89 years are found 
(see Figure 5).  Apart from the long-term periods, periods of $p_1= 0.981(V_m^2 = 0.857)$,$p_2=1.952(V_m^2 = 0.91)$, and $p_3=2.935(V_m^2 = 0.873)$ years
 have
 also been found. They can be considered as real periods as discussed above.  It is clear that these periods are correlated as $p_1 
\approx 1 yr$, $p_2 \approx2p_1$,$p_3 \approx 3p_1$. The longer ones are likely the harmonics of the shortest one, $p_1 \approx 1 yr$. 
The period of one year maybe result from the effect of the Sun on the observations. There is also a signature of a period of about 7.5 years
in Figure 5, which we do not take  as  real as it is not so conspicuous as compared with the ``flat'' noise.  
When only the post-1970 data (see Figure 2c) are used, periods of 0.08, 0.60 and 0.89 year are found
 (see Figure 6). Also,  from figure 6, it can be seen that there is a broad minimum, which corresponds to a period of $7.5 \pm 1.3$ 
years, but its $f=0.16$ suggests that it is a weak period; for other minima, they are not conspicuous as compared with the noise, we do
 not consider them as signature of periods.

\begin{figure}
\epsfxsize=12cm
$$
\epsfbox{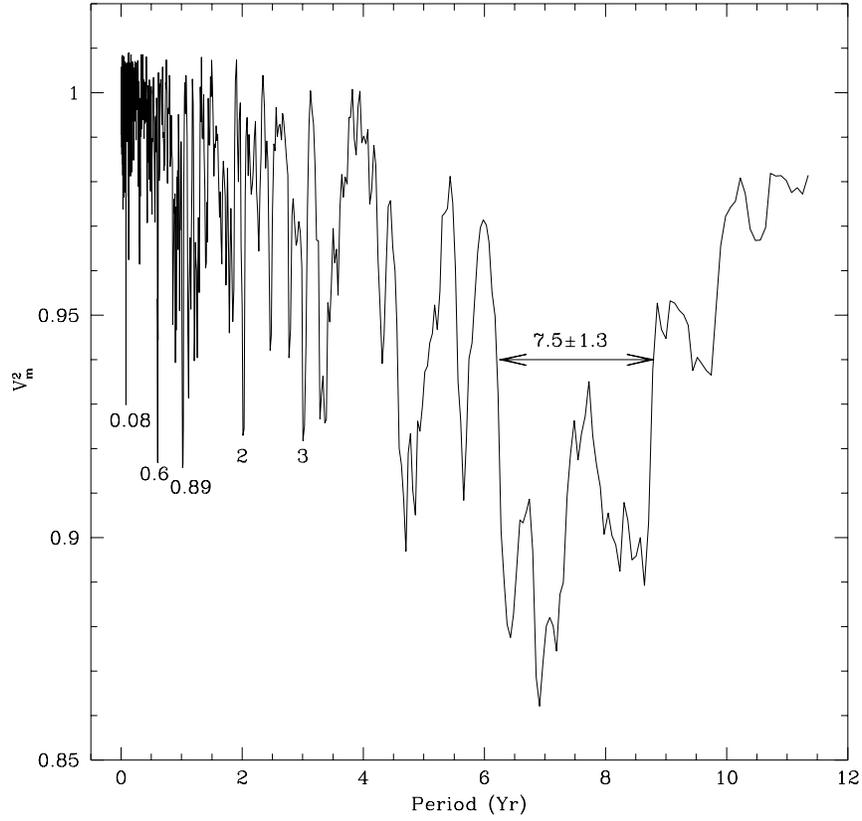}
$$
\caption{ Diagram of $V_{m}^2$ against the trial periods for data covering the observation period  1970 - 1996 
($m$=5 is adopted). The broad minimum corresponds to a period of $7.5\pm 1.3$ years and others to periods of $0.08$, $0.60$, 
$ 0.89 $, $2$ and $3$ years respectively. }
\end{figure}

\section{Discussion}

BL Lacertae is the archetype of the BL Lac objects and one of the best studied objects.  Its spectrum is  usually featureless, but weak 
emission lines are indeed identified when the source is in the fainter state(Vermeulen et al. 1995, Corbett et al. 1996a,b).  
From quasi-simultanous photometry, Barbieri et al. (1988) found that there was no correlation between the
 continuum brightness and the strength of the spectral lines.  The optical variability is extremely irregular over periods of hundreds of
 days of continuous observation (Bertaud et al. 1969; Du Puy et al. 1969; Racine 1970; Tritton \& Brett 1970).  Shen \& Usher (1970) had 
shown historic light curves back to as early as 1896.  Two large bursts can be seen to have occurred in 1928/1929 and 1954 (see Figure
 1 of their paper).  The brightness reached $m_{pg}$=12.4 during the 1928/1929 burst.  Shen \& Usher proposed that the object would undergo
 a long-term brightness increase and decline spanning 10 to 15 years.  In 1988, Webb et al. compiled post-1970 optical observations
, investigated the variation period, and found periods of 0.31, 0.60, and 0.88 year. In 1996, Markenko et al. also investigated the
long-term period from the light curve covering the period of 1968 to 1989 and found a period of  $7.8\pm0.2$ years and a possible period 
of 34 years.  In addition, some rapid variability over short time scales has been reported: for example, a variation of 1.5 magnitude
 over a time scale of 20 hours (Weistrop, 1973), daily variation as great as 0.3 magnitude (Carswell et al. 1974),  variation of
 0.56 magnitudes over a short time scale of 40 minutes in the B band (Xie et al. 1988b, 1990), and variation of 0.1 magnitude over 
30 minutes in V band(Corbett et al. 1996a).

\subsection{Periodicity}

To establish the reality of periods, we need a duration of data. The length of the data record required to demonstrate
a periodicity depends on signal-to-noise, systematic errors, regularity in times of measurements, and the nature of the 
underlying variation. Perfect data on a perfectly periodic variable might not
require more than 1.5 periods to yield an accurate measure of the period 
. But for some objects whose existing data do not sample
their light curves at regular intervals,  a much longer time span of data can be used to demonstrate that any possible period is not simply a random event and probably has some physical significance (Kidger et al. 1992 
also see Fan et al. 1997).

For BL Lacertae, the data are not distributed equally during the period of 1896-1996.  When the periodicity analysis method 
(Jurkevich et al. 1971) is used to deal with the 5-day averaged B data, the following periods (see Table 2) are found ($m=10$ is adopted):

\begin{table}
\caption[]{Periodicity derived from the B Light Curve}
\begin{tabular}{ccc} \hline
$P_A$=13.97 $\pm$ 0.75(yr)  &  $V_m^2$=0.825  &  f=0.21 \\ 
$P_B$=19.55 $\pm$ 0.40(yr)  &  $V_m^2$=0.784  &  f=0.28 \\ 
$P_C$=29.16 $\pm$ 0.50(yr)  &  $V_m^2$=0.769  &  f=0.30 \\ 
$P_D$=35.00 $\pm$ 2.00(yr)  &  $V_m^2$=0.774  &  f=0.29 \\ 
$P_E$=38.89 $\pm$ 0.40(yr)  &  $V_m^2$=0.677  &  f=0.48 \\ \hline
\end{tabular}
\end{table}

  As introduced in section 3.1, $P_B, P_C, P_D, P_E$ are possible periods as their values of $f$'s are greater than 0.25 but less 
than 0.50, and $ P_A $ is a weak one as 
its $f$ is less than 0.25. Because the unequally distributed data cover
a period of 100 years, the derived longer periods ($P_B, P_C, P_D, P_E$) should be confirmed with more observations as argued in the paper of Kidger et
al (1992).  For the 13.97-year period, it is clear that the data sample is seven times the length of the period. 
From figure 5 we can see that the minimum corresponding to this period is also
conspicuous as compared with the ``noise'', so it can be considered as a real period even though its  $f$ being less than 0.25. 
This  period is in agreement with that proposed by Shen \& Usher (1970).    
Besides, from figures 5 and 6, we know that there are other periods(see section 3.2), which can be considered as real periods 
as argued in section 3.1
 although their
 $f$'s are less than 0.25.  We think that the period of  $p_1\approx$ 1 yr is caused by the Sun, interrupting the observations while the 
period of 0.08 years is likely to be caused by the Moon in the same way.

\subsection{Comparison with previous results}
BL Lacertae has been investigated for long-term periods in papers of Webb et al.(1988), Marchenko et al.(1996) and in the 
present one.  Webb et al. (1988) used  Fourier analysis to search for periodicity from the data covering the period of 1970 through 
1986; the periods of 0.60 and 0.88 year reported in their paper also appear in our analysis, but the period of 0.31 years found by 
Webb et al.(1988) is not found.  One of the reasons may be from the fact that their method is not good enough to deal
 with the unequally spaced data series.  Marchenko et al.(1996) used the method called ``whitening'' of time
 series to investigate the periods from data covering an observation period from 1968 to 1989 (Marchenko et al. 1996), this method
 has the same advantage for finding periods in  unequally spaced time series as the method used in the present paper as compared 
with the Fourier method.  But the method used in the present paper is  simpler for computation as compared with the ``whitening'' method. 
The period of $7.5$ 
years appearing in Figure 6 is consistent with the period of $7.8$ found by Marchenko et al.(1996). But it is only a very weak period 
in our discussion, which may come from the fact that the post-1970 data sample is only about four times the length of the period. 
The period of $35\pm2.0$ years found in our analysis is in good agreement with the period of $34$ years found by Marchenko et al.(1996),
 but it is also a possible one in our analysis; its reality should be
 discussed 
with more observation in the future.

\acknowledgements The authors thank an anonymous referee for the valuable comments and G. Adam for the critical reading. JHF
thanks Dr. G. Zhao's  invitation for a short-period to visit Young Astronomer
Center, Beijing Astronomical Observatory so that he can do part work of the paper.
This work is supported by the National Scientific Foundation of China(the NinthFive-year Important Project) and the National 
Pandeng Project of China.

\cite{}

{}


\begin{thebibliography}{}

\bibitem[]{} Arimoto J. Sadakano K., Honda, S; \& Tanake, K. 1997, PASP, 109, 300
\bibitem[]{} Bai, J.M., Xie, G.Z., \& Li, K.H. et al. 1998 (in preparation)
\bibitem[]{} Barbieri C. Cristiani, S. \& Romano, C. 1982,AJ, 87, 616
\bibitem[]{} Barbieri C. Cappellaro, E. Romano, G. Turatto, M, Szuszkiewicz, E. 1988, A\&AS,76, 477 
\bibitem[]{} Barbieri C. Romano, G. \& Zambon, M. 1978, APJS, 31, 401
\bibitem[]{} Belokon', E.T.,  1987, Astrofizika, vol 27, 429
\bibitem[]{} Bertaud G. Wlerick, G. Veron, P. Dumortier, B., Bigay, J., Paturei, G., Duruy, M., Savesky, P.,  1973, A\&A, 24, 357
\bibitem[]{} Bertaud, G. Dumortier, B. Veron, P. Wlerick, G., Adam, G., Bigay, J., Gaarnier, R., \& Duruy, M. 1969 A\&A, 3, 436
\bibitem[]{} Blandford, R.D. 1996, in 'Blazar Continuum Variability' 
  eds. H.R. Miller, J.R. Webb, and J.C. Noble, ASP Conf. Series VoL. 110, 475
\bibitem[]{} Bregman, J.N. Glassgold, A.E., Huggins, P.J. Neugebauer, G., Soifer, B.T., Matthews, K., Elias, J.H., Webb, J.R., 
Pollock, J.T., Leacock, R.J., Smith, A.G., Aller, D., Aller, M.F., Hughes, P.A., Maccagni, D., Garilli, B., Giommi, P., Miller, J.S., 
Stephens, S., Balonek, T.J., Dent, W.A., Kinsel, W., Winniewski, W.Z., Williams, P.M., Brand, P.W.J.L., Ku, W.H.-H.1990, ApJ, 352, 574
\bibitem[]{} Cannon, R.D. Penston, M.V., Brett, R.A., 1971, MNRAS, 152, 79
\bibitem[]{} Carini, M.T. Miller, H.R., Noble, J.C., \& Goodrich, B.D., 1992, AJ, 104, 15
\bibitem[]{} Carswell, R.F., Strittmatter, P.A. Williams, R.E., Kinman, T.D., \& Serkowski, K., 1974, ApJ, 190, L101
\bibitem[]{} Catanese, M., Akerlof, C.W. Biller, S.D., Boyle, P., Buckley, J.H., Carterlewis, D.A., Cawley, M.F., Connaughton, V., 
Dingus, B.L., Fegan, D.J., Fichtel, C.E., Finley, J.P., Gaidos, J.A., Gear, W.K., Hartman, R.C., Hillas, A.M., Krennrich, F., 
Lamb, R.C., Lessard, R.W., Lin, Y.C., Mcenery, J.E., Maracher, F., Mohanty, G., Mukherjee, R., Quinn, J., Robson, E.I.,
 Rudgers, A.J., Rose, H.J., Samuelson, F.W., Sembrosk, G., Schubnell, M.S., Stevens, J.A., Teraesranta, H., Thompson, D.J.,
 Weekes, T.C., Wilson, C., Zweerink, J. ;1997, ApJ, 480, 562
\bibitem[]{} Corbett, E. Robinson, Axon, D.J. Hough, J. Jeffries, R.A. Thurston, M.R., Young, S.  1996a, MNRAS, 281, 737
\bibitem[]{} Corbett, E. Robinson, A. Hough, J. Young, S. 1996b, in 'Blazar Continuum Variability' 
  eds. H.R. Miller, J.R. Webb, and J.C. Noble, ASP Conf. Series VoL. 110, 160
\bibitem[]{} Corsco, G.J. Schultz, J. Dey, A. 1986, PASP, 98, 1287
\bibitem[]{} Du Puy, D. Schmitt, J.L. McClure, R. van den Bergh, S. \& Racine, R.  1969, ApJ, 156, L235
\bibitem[]{} Fan J.H. \& Su C.Y. 1997, Acta Astronomica Sinica (in press)
\bibitem[]{} Fan J.H. Xie, G.Z;, Lin, R.G. et al. 1997, A\&AS, 125, 525
\bibitem[]{} Fan J.H., Xie, G.Z., \& Wen S.L. 1996, A\&AS, 116, 409
\bibitem[]{} Hagen-Thorn, V.A. Marchenko, S.G., Mikolaichuk, O.V. Yakovleva, V.A. 1997, Astronomy Report, 41, 1
\bibitem[]{} Hagen-Thorn, V.A. Marchenko, S.G., Yakovleva, V.A. 1984, Sov. Astron 28(5), 538
\bibitem[]{} Jurkevich, I. 1971, Ap\&SS, 13, 154
\bibitem[]{} Jurkevich, I., Usher, P.D., Shen, B.S.P. 1970, Ap\&SS, 10, 402
\bibitem[]{} Kawai, N. Matsuoka, M. Bregman, J.N., Aller, H.D., Aller, M.F., Hughes, P.A., Balbus, S.A., Balonek, T.J., 
Chambers, K.S., Clegg, R.E.S., Clements, S.D., Leacock, R.J., Smith, A.G., Goodrich, R., Miller, J.S., Hereld, M., Hoare, M.G.,
 Hughes, V.A., Miley, G.K., Moriarty-Schieven, G.H., Matthews, C., Neugebauer, G., Ohashi, T., Roche, P.F., Thronson, H.A., 
Valtaoja, E., Terasranta, H., Webb, J.R., Wills, B.J., Wills, D. 1991, ApJ, 382, 508
\bibitem[]{} Kerrick, A.D. Akeriof, C.W., Biller, S., Buckley, I., Carter-Lewis, D.A., Cawley, M.F., Chaniell, M., Connaughton, V., 
Fegan, D.J., Fennell, S., Gaidos, J., Hillas, A.M., Kwok, P.W., Lamb, R.C., Lappin, T., Lessard, R., Mcenery, J., Meyer, D.I., 
Mohanty, G., Quinn, J., Rose, H.J., Rovero, A.C., Semberoski, G., Schubnell, M.S., Punch, M., Weeks,T.C., West, M., Wilson, C., 
Zweerink, J. 1995, ApJ, 452, 588
\bibitem[]{} Kidger, M.R, Takalo, L. Sillanpaa, A. 1992, A\&A, 264, 32
\bibitem[]{} Kidger, M.R, Takalo, L. 1990, A\&A, 239, L9
\bibitem[]{} Kidger, M.R; 1989, A\&A, 226, 9
\bibitem[]{} Kidger, M.R, 1988, PASP, 100, 1248
\bibitem[]{} Liu, F.K. Xie, G.Z., Bai J.M. 1995, A\&A, 295, 1
\bibitem[]{} Lu P.K. 1972, AJ, 77, 829
\bibitem[]{} Maesano, M., Montagni, F., Massaro, E., Nesci, R. 1997, A\&AS, 122, 267
\bibitem[]{} Marchenko, S.G., Hagen-Thorn, V.A., Yakovleva, V.A. Mikolaichuk, O.V. 1996 in 'Blazar Continuum Variability' 
  eds. H.R. Miller, J.R. Webb, and J.C. Noble, ASP Conf. Series Vol. 110, 105
\bibitem[]{} Mead, A.R.G., Ballard, K.R., Brand, P.W.J.L. Hough, J.H., Brindle, C., Bailey, J.A. 1990, A\&AS, 83, 183
\bibitem[]{} Miller, J.S. French, H.B., Hawley, S.A. 1978, ApJ, 219, L85
\bibitem[]{} Miller, H.R. \& McGimsey B. 1978, ApJ, 220, 19
\bibitem[]{} Moles, M., Garcia-Pelayo, J.M. Masegosa, J., Aparicio, A.  1985, ApJS, 58, 255
\bibitem[]{} Mutel, R.L. Phillips, R.B. 1987, in Superluminal Radio Sources, eds, J.A. Zensus \& T.J. Pearson,(Cambridge), p60
\bibitem[]{} O'Dell, S.L. Pushell, J.J., Stein, W.A., Warner, J.W;, 1978, ApJS, 38, 267
\bibitem[]{} Okyudo, M. 1993, Annu. Rep; Nishi-Harima, Obs. 3. 1
\bibitem[]{} Puschell, J.J. \& Stein, W.A. 1980, ApJ, 237, 331
\bibitem[]{} Qogun Y. A. et al 1994, Astro. Journal Letter of Russian (in Russian) 20, 266
\bibitem[]{} Quinn, J. et al. 1995, in Proc. 24th Int. Cosmic Ray Conf;(Room), 2, 269
\bibitem[]{} Racine, R. 1970, ApJ, 159, L99
\bibitem[]{} Shen B.S.P. \& Usher, P.D. 1970, Nat. 228, 1070
\bibitem[]{} Sillanpaa, A. Takalo, L.O., Pursimo, T., Lehto, H.J., Nilsson, K., Heinamaki, P., Kidger, M.R., De Ddiego, J.A., 
Boltwood, P., Dultzin-Hacyan, D., Benitez, E., Turner, G.W., Robertson, J.W., Honeycut, R.K., Efimov, Yu. S., Shakhovskoy, N., 
Schramm, K., Borgeest, U., Linde, J.V., Weneit, W., Kuehl, D., Schramm, T., Sadun, AA., Grashuis, R., Heidt, J., Wagner, S., 
Bock, H., Kuemmel, M., Heines, AA., Fiorucci, M., Tosti, G., Ghisellini, G., Roiteri, C.M., Villata, M., De Francesco, G.,
 Lanteri, L.,1996a, A\&A, 305, L17
\bibitem[]{} Sillanpaa, A.  Takalo, L.O., Pursimo, T., Nilsson, K., Heinamaki, P., Katajainen, S., Pietila, H., Hanski, M.,
 Rekola, R., Kidger, M.R., Boltwood, P., Turner, G.W., Robertson, J.W., Honeycut, R.K., Efimov, Yu. S., Shakhovskoy, N.,
 Fiorucci, M., Tosti, G., Ghisellini, G., Roiteri, C.M., Villata, M., De Francesco, G., Lanteri, L., Chiabergo, M., Peila, A.,
 \& Heidt, J. 1996b,  A\&A, 315, L13
\bibitem[]{} Sillanpaa, A. Mikkola, S. Valtaoja, L.  1991, A\&AS, 88, 225
\bibitem[]{} Sillanpaa, A. Haarala, S. Valtonen, M.J. Sundelius, B., \& Byrd, G.G. 1988a, ApJ, 325, 628
\bibitem[]{} Sillanpaa, A. Haarala, S. Korhonen, T. 1988b, A\&AS, 72, 347
\bibitem[]{} Sitko, M.L. \& Sitko, A.K.  1991, PASP, 103, 160
\bibitem[]{} Sitko, M.L., Schmidt, G.D, Stein, W.A. 1985, ApJS, 59, 323
\bibitem[]{} Sitko, M.L. Stein, W.A., Zhang, Y.X., Wisniewski, W.Z., 1983, PASP, 95, 724
\bibitem[]{} Smith, P.S. Balonek, T., Elston, R. Heckert, P.A; 1987, ApJS, 64, 459
\bibitem[]{} Smith, P.S. Balonek, T., Heckert P. Elston, R. Schmidt, G.D.; 1985, AJ, 90, 1184
\bibitem[]{} Takalo, L.O.  1991, A\&AS, 90, 161
\bibitem[]{} Tritton, K.P. \& Brett, R.A. 1970, Observatory, 90, 110
\bibitem[]{} Vermeulen, R.C., Ogle, P.M. Tran, H.D., Brobne, I.W.A., Cohen, M.H., Readhead, A.C.S., Taylor, G.B., Goodrich, R.W.
 1995, ApJ, 452, L5
\bibitem[]{} Vermeulen, R.C. \& Cohen, M.H. 1994, ApJ, 430, 467
\bibitem[]{} Visvanathan, N. 1973, ApJ, 179, 1
\bibitem[]{} Webb, J.R., Smith, A.G., Leacock, R.J. Fitzgibbons, G.L., Gombola, P.P., \& Sheppard, D.W.,  1988, AJ, 95, 374
\bibitem[]{} Weistrop, D, 1973, Nat, 241, 157
\bibitem[]{} Xie G.Z.   Li, K.H., Zhang, Y.H., Liu, F.K., Fan, J.H., \& Wang, J.C.  1994, A\&AS, 106, 361
\bibitem[]{} Xie G.Z. Li, K.H., Liu, F.K., Lu, R.W., Xu, J.X., Fan, J.H., Zhu, Y.Y., \& Cheng, F.Z.  1992, ApJS, 80, 683
\bibitem[]{} Xie G.Z. Li, K.H., Cheng, F.Z., Hao, P.J., Li, Z.L., Lu, R.W., \& Li, G.H. 1990, A\&A, 229, 329
\bibitem[]{} Xie G.Z. Li, K., Zhou, Y., Lu, R.W., Wang, J.C., 1988a, AJ, 96, 24
\bibitem[]{} Xie G.Z. Lu, R.W., Zhou, Y., Hao, P.J., Zhang, Y., Li, X.Y., Liu, X.D., \& Wu, J.X. 1988b, A\&AS, 72, 163
\bibitem[]{} Xie G.Z. Li, K.H., Bao, M.X., Hao, P.J., Zhou, Y., Liu, X.D., \& Deng, L.W. 1987, A\&AS,67, 17
\end{thebibliography}
\end{document}